\documentclass[jmp]{revtex4-1}
\usepackage{bm}
\usepackage{dcolumn}
\usepackage {amsfonts,amsmath}
\begin{document}

\title {A modular-invariant modified Weierstrass sigma-function as a
  building block for lowest-Landau-level wavefunctions on the torus}

\author{ F. D. M. Haldane}
\email{haldane@princeton.edu} 
\affiliation{Department of Physics, Princeton
  University, Princeton NJ 08544-0708, USA}

\date{June 3, 2018, revised June 20, 2018}
\begin{abstract}
A ``modified'' variant of the Weierstrass sigma, zeta, and elliptic
functions is proposed whereby the zeta function is redefined by  $\zeta(z)$ $\mapsto$
$\widetilde \zeta(z)$ $\equiv$ $\zeta(z) - \gamma_2z$, where $\gamma_2$ is
a lattice invariant
related to  the almost-holomorphic modular invariant of the
quasi-modular-invariant weight-2 Eisenstein series.  If $\omega_i$ is
a primitive half-period, $\widetilde\zeta(\omega_i)$ = $\pi \omega_i^*/A$,
where $A$ is the area of the primitive cell of the lattice.   The
quasiperiodicity of the modified sigma function is much simpler than
that of the original, and it becomes the building-block for the
modular-invariant formulation of lowest-Landau-level wavefunctions on
the torus.   It is suggested that the ``modified'' sigma function is
more natural than the original Weierstrass form, which was formulated
before quasi-modular forms were understood.     For the high-symmetry
(square and hexagonal) lattices, the modified and original sigma
functions coincide. 
 \end{abstract}
\maketitle

%\section {A  ``modified''  Weierstrass sigma function}

In 1862, Weierstrass\cite{weier}, as part of his work on elliptic
functions,
introduced an odd meromorphic ``zeta function''
$\zeta(z;\Lambda)$ which has simple poles with residue 1 on a lattice $\Lambda$ =  $\{L\}$ = 
$\{2m\omega_1 + 2n\omega_2\}$ in the complex plane, where $m$ and
$n$ are integers: $m,n$ $\in$ $\mathbb Z$.
Using the metric $d(z_1,z_2)$ = $|z_1-z_2|$, the lattice
has a primitive cell of area
\begin{equation}
A(\Lambda) = 2|\omega_1^*\omega_2 - \omega^*_2\omega_1|  > 0.
\end{equation}
Here $\omega_i$ will mean any primitive half-period of the lattice, so
$2\omega_i$ $\in$ $\Lambda$, but
$2\omega_i/p$ $\not \in \Lambda$ for any integer $p > 1$.
A pair of primitive half-periods  $(\omega_1,\omega_2)$  is a basis of
$\Lambda$ if
$\omega_1 + p\omega_2$ and $\omega_2 + p\omega_1$ are primitive for
all integers $p$.   Once a basis is chosen, it is useful to define
$\omega_3$ =$-(\omega_1 + \omega_2)$, which is also a primitive half-period.

The Weierstrass  zeta function is defined by the convergent sum
\begin{equation}
\zeta(z;\Lambda)
= \frac{1}{z} + {\sum_{L\ne 0}}' \left ( \frac{1}{z-L}  + \frac{1}{L} 
  + \frac{z}{L^2}\right ) = \frac{1}{z} + {\sum_{L\ne 0}}'
\frac{z^3}{L^2(z^2-L^2)}, 
\label{zsum}
\end{equation}
where the primed sum is over all $L\in \Lambda$ except $L = 0$.
Near $z$ = 0, it has a Laurent expansion
\begin{equation}
\zeta (z;\Lambda) = \frac{1}{z} - \sum_{k=2}^{\infty} \gamma_{2k}(\Lambda)
z^{2k-1},
\label{laurent}
\end{equation}
where
$\gamma_{2k}(\Lambda)$ are lattice invariants 
\begin{equation}
\gamma_{2k}(\Lambda) = {\sum_{ L \ne 0}}' \frac{1}{L^{2k}}, \quad k \ge 2,
\label{mod} 
\end{equation}
which are related to the Eisenstein series $G_{2k}(\omega_2/\omega_1)$
= $ (2\omega_1)^2\gamma_{2k}(\Lambda)$.

The Weierstrass elliptic function $\wp(z;\Lambda)$  is (minus) the
derivative of the zeta function, and is an even doubly-periodic meromorphic
function of $z$ defined by
\begin{equation}
 \wp(z;\Lambda)  = -\zeta'(z;\Lambda) \equiv
                   -\frac{\partial}{\partial z}\zeta(z;\Lambda)
= \frac{1}{z^2} + {\sum_{L\ne 0}}' \left (\frac{1}{(z- L)^2} -
  \frac{1}{L^2}\right ).
\end{equation}
The  elliptic function seem to have been Weierstrass' main interest,
but the emphasis here will be  on  an application of the  quasiperiodic
 sigma function, defined in terms of the zeta function by
\begin{equation}
\sigma (z;\Lambda) = \lim_{\epsilon\rightarrow 0} \epsilon
\exp \left ( \int_{\epsilon}^z dz' \zeta (z';\Lambda) \right ),
\label{sigmadef}
\end{equation}
where the integral is independent of the path because the poles of the
zeta function have unit residue.   The sigma function is an odd holomorphic
function with  only simple  zeroes which are at the lattice points
$z$ $\in$ $\Lambda$.   Then
\begin{equation}
\zeta(z;\Lambda) = \frac{\sigma'(z;\Lambda)}{\sigma(z;\Lambda)},
\quad  \sigma'(0;\Lambda) = 1.
\end{equation}

A curious feature  of the Weierstrass definitions
is the absence of a term 
$O(z)$ in the Laurent expansion (\ref{laurent})  for $\zeta(z;\Lambda)$.  
Weierstrass  presumably made
this choice because the definition (\ref{mod}) of $\gamma_{2k}(\Lambda)$ is ill-defined for $2k$
= 2, and he
was working long before Ramanujan investigated quasi-modular forms.
However, based on Ramanujan's work\cite{ram} on such forms,  there is a 
``natural'' definition of $\gamma_2(\Lambda)$ given by ``Eisenstein summation''
\begin{equation}
\Gamma_2(\omega_1,\Lambda) \equiv\sum_{m\ne 0} \frac{1}{(2m\omega_1)^2} +
\sum_{n\ne 0}  \left (\sum_{m} 
\frac{1}{(2m\omega_1 + 2n\omega_2)^2} \right )
\label{Gam} 
= \gamma_2(\Lambda) + 
\frac{\pi}{A(\Lambda)}\frac{\omega_1^*}{\omega_1},
\end{equation}
where a sum over $n$ is interpreted as the limit as $N$
$\rightarrow$ $\infty$ of a sum over $n$ in the interval $[-N,N]$.
$\Gamma_2(\omega_i,\Lambda)$ is related to the weight-2
quasi-modular-invariant
of the $2k$ = 2 Eisenstein series, and
 $\gamma_2(\Lambda)$ is a lattice invariant related to the corresponding
 ``almost-holomorphic modular form''.

In fact knowledge of the invariant $\gamma_2(\Lambda)$ 
had already been available since 1847, fifteen years before
Weierstrass presented his work, in the 
published (but apparently ignored) work of
Eisenstein\cite{eis} who died in 1851.     This work was later briefly brought to
attention in 1891 in a lecture by Kronecker, who  appears to have
belatedly recognized its
importance, but who unfortunately died just fifteen days after the lecture,
leaving extensive notes that remained unpublished, and which  were  believed lost until
recently rediscovered\cite{ems}.   
Some of Eisenstein's results remained overlooked and forgotten 
until they were reviewed in detail and publicized in a 1976  book by Weil\cite{weil}.

Then  defining ``modified'' Weierstrass functions by including the
$2k$ = $2$ term in the Laurent-series definition of
$\zeta(z;\Lambda)$ has the effect
\begin{align}
\wp(z,\Lambda) &\mapsto \widetilde \wp(z,\Lambda)  = \wp(z,\Lambda)+ \gamma_2(\Lambda), \\
\zeta(z,\Lambda) &\mapsto \widetilde \zeta(z,\Lambda) = 
\zeta(z,\Lambda) - \gamma_2(\Lambda)z, \\
\sigma(z,\Lambda) &\mapsto \widetilde \sigma(z,\Lambda) =
e^{-\frac{1}{2}\gamma_2(\Lambda)z^2}\sigma(z,\Lambda).
\end{align}
Note that $\gamma_2$ vanishes for both the hexagonal  lattice ($\gamma_4$ = 0),
and the square lattice ($\gamma_6$ = 0), so in these high-symmetry
lattices, the ``modified'' and ``original'' Weierstrass functions
coincide.     Specification of the pair of complex invariants $(\gamma_4,\gamma_6)$
(so long as they do not both vanish) uniquely parametrizes the lattice $\Lambda$.
For ease of notation, the dependence of these functions on the lattice
$\Lambda$ will now be left implicit, unless needed.

An important property of both the original and the modified Weierstrass
functions is that they depend only the lattice $\Lambda$, and do not
depend on a particular choice of basis $(\omega_1,\omega_2)$ (or on a
choice of orientation of the basis).   The group $SL(2,\mathbb Z)$ of
(orientation-preserving) changes of basis is known as the modular group,
and (with a slight abuse of terminology)  the basis-independence of
the Weierstrass functions may be called ``modular-invariance''.

The ``modified'' forms of the Weierstrass functions were 
in fact already present  in Eisenstein's 
work\cite{eis}, as reviewed by Weil\cite{weil}.  In the
notation
used by Weil, $\widetilde \zeta(z)$ is ``$E_1(z)$'',   $\widetilde \sigma (z)$ is
``$\varphi (z)$'', and $\widetilde \wp (z)$ is ``$E_2(z)$''. 
A key result also found  in Eisenstein's work\cite{eis} is that the
non-meromorphic function
\begin{equation}
\widehat \zeta(z,z^*)  =     \widetilde \zeta(z) - \frac{\pi z^*}{A}
\label{eisfun}
\end{equation}
is a doubly-periodic modular-invariant 
function (referred to as ``$E_2^*(z)$'' by Weil\cite{weil}).

In recent work\cite{recent}, there is renewed interest in this
function, which  has been called
``Eisenstein's (periodic) completion'' of the
Weierstrass zeta function.
Generically, it has a lattice of  simple poles
at $z$ $\in$ $\{L\}$, three lattices of 
simple holomorphic zeroes
and two lattices of simple antiholomorphic zeroes.
Each lattice of holomorphic zeroes is invariant under inversion about
a pole, and they
are located at the points
$z$ $\in$ $\{ L + \omega_i\}$,  as a consequence of the relation
\begin{equation}
\zeta (\omega_i)  \equiv \eta_i = \gamma_2 \omega_i 
+ \frac{\pi \omega_i^*}{A}; \quad \widetilde
\zeta(\omega_i) = \frac{\pi \omega_i^*}{A}.
\label{eta1}
\end{equation}
The lattices of antiholomorphic zeroes are mapped into each other by
the inversion.
 The six antiholomorphic zeroes
 which are closest to a pole define the corners of a
 six-sided unit cell centered on the pole, with holomorphic zeroes at
 the centers of the sides.   (This is similar to a ``Wigner-Seitz'' or
 Voronoi cell, but direct inspection using the Weierstrass functions
implemented in Mathematica\textsuperscript{\textregistered}   shows that
the locations of the antiholomorphic zeroes are not in general given
by the Voronoi construction using the metric $d(z,z')$ = $|z-z'|$.)
In the limit of a simple-rectangular lattice, one pair of sides shrink
to points where two antiholomorphic zeroes combine with a holomorphic
one, leaving a rectangular unit cell with a simple antiholomorphic
zero at its corners.
Eisenstein's periodic function (\ref{eisfun}) is a sum of 
``holomorphic'' (meromorphic) and antiholomorphic parts:
the ``modified zeta function'' is just the ``holomorphic'' part of
Eisenstein's periodic completion of the zeta function.

A consequence of the relation (\ref{eta1}) is the well-known property
\begin{equation}
\eta_1\omega_2 - \eta_2\omega_1 =  \frac{\pi
  (\omega_1^*\omega_2-\omega_1^*\omega_2)}{A} =
\pm {\textstyle\frac{1}{2}}i\pi ,
\end{equation}
which is an orientation-independent version of a standard relation
between the $\eta_i$ and $\omega_i$. (Note that the formulation presented here 
does \textit{not} require the  orientation of the basis to be chosen so
that  the imaginary part of $\omega_2/\omega_1$ (or ``$\tau$'')  is
positive.)     The explicit expression (\ref{eta1}) involving
$\gamma_2(\Lambda)$
does not seem to be
as widely known in the literature on Weierstrass functions (it does
not seem to appear in the standard reference sources), but is
easily verified by
using a  basis
$(\omega_1,\omega_2)$, with $\omega_1$ = $\omega_i$,  
and  evaluating the lattice sum (\ref{zsum}) in the order used
in the definition (\ref{Gam}) of $\Gamma_2(\omega_1;\Lambda)$:
\begin{equation}
\eta_1= 
S(\omega_1,\omega_2)
+ \omega_1 \Gamma_2(\omega_1,\Lambda)  ,
\end{equation}
where the sum over $1/L$ is dropped because it vanishes by symmetry,
and 
\begin{align}
S(\omega_1,\omega_2)  
&= \sum_{n} \left ( \sum_m \frac{ 1}{(2m+1)\omega_1 +
  2n\omega_2}\right )
= -
\frac{\pi}{2\omega_1}
\sum_n \tan (n\pi \omega_2/\omega_1) = 0. 
\end{align}
The terms in the sum $S$ are odd under $n \mapsto  -n$, 
so it also vanishes.  (This somewhat-heuristic symmetry-based argument can
be confirmed  by more rigorous ``$\epsilon \rightarrow 0$'' treatments.)

The values of $\eta_i$ $\equiv$   $\zeta(\omega_i;\Lambda)$ 
appear as a complication in
the expressions for the quasiperiodicity of the Weierstrass sigma function.
The  equivalent quantity in the 
``modified'' sigma function has the  much simpler form
\begin{equation}
\widetilde \eta_i \equiv  \eta_i - \gamma_2\omega_i =  \pi
\omega_i^*/A,
\label{new}
\end{equation}
so the modified values of the ``strange numbers'' $\eta_i$ that permeate various
sigma-function identities become  the natural primitive ``reciprocal half-lattice
vectors''
conjugate to  the primitive ``half-lattice vectors'' $\omega_i$.
All the various  Weierstrass sigma-function identities involving $\eta_i$ remain true
when the Weierstrass functions are replaced by their ``modified''
forms and $\eta_i$ is replaced by $\pi \omega_i^*/A$.
In particular, for $q$ = $\exp (i \pi \tau)$ with $\tau$ = $ \omega_2/\omega_1$,
$\sigma(z;\Lambda)$ has a rapidly-converging product representation
\begin{equation}
\sigma(z) =  z e^{\eta_1z^2/2\omega_1}\left (
\frac{\sin u}{u} \prod_{n=1}^{\infty} 
\frac{\left ( q^{2n} + q^{-2n}  - 2\cos 2u \right )}{(q^n-q^{-n})^{2}}\right )
\equiv
z e^{\eta_1z^2/2\omega_1}
\left (\frac{\vartheta_1(u|\tau)}{u\vartheta_1'(0|\tau)}\right ), \quad
  u = \frac{\pi z}{2\omega_1},
\label{full}
\end{equation}
where $\vartheta_1(u|\tau)$ is the Jacobi theta-function with zeroes
at $\{m\pi  +  n\pi \tau\}$. 
The modified form $\widetilde
\sigma(z)$ is given by the same formula, with the replacement
$\eta_1 \mapsto \widetilde \eta_1$ = $\pi\omega_1^*/A$.
Note that while the constructive definition of the original Weierstrass 
sigma function (as an infinite product) 
follows from the lattice sum (\ref{zsum}), the constructive
definition of the modified sigma function is directly given by
(\ref{full}) with
$\eta_1 \mapsto \pi \omega^*_1/A$.
The expression (\ref{full}) is  not a constructive definition for the
original Weierstrass sigma function, as it requires a separate evaluation of
$\gamma_2(\Lambda)$ in order to provide $\eta_1$.    Similarly, an
expression based on (\ref{zsum}) does not provide a constructive
definition of the modified form, because it would then also require
a separate evaluation
of $\gamma_2(\Lambda)$.

The quasiperiodicity of the modified sigma function now has
the more ``elegant'' form
\begin{equation}
\widetilde \sigma(z + L)
= \xi(L) e^{(\pi L^*/A)(z + \frac{1}{2}L)}\widetilde\sigma (z),
\end{equation}
where $\xi(L)$ is the \textit{parity} of $L$: 
$\xi(L) = 1$  if $\frac{1}{2}L \in \Lambda$, and $-1$ otherwise.
In the limit $|z| \rightarrow 0$, the holomorphic sigma functions (both the
modified and original forms) have the property
\begin{equation}
\lim_{z\rightarrow 0} \frac{\sigma(z)}{z} \rightarrow 1.
\end{equation}
It is worth pointing out that the non-holomorphic function
\begin{equation}
\mathcal Z(z,z^*) \equiv  \widetilde \sigma(z) e^{-\frac{1}{2}\pi z^*z/A}
\end{equation}
has the property that its absolute value  $|\mathcal Z(z,z^*)|$ is a function with the
lattice periodicity.    This may be regarded as the nearest equivalent
to the linear function $f(z)$ = $z$ that can be made (quasi)periodic,
with
\begin{equation}
\lim_{z\rightarrow 0} \frac{\mathcal Z(L + z,L^* +  z^*)}{z}
\rightarrow \xi (L).
\end{equation}
This function is essentially an analog of the  sigma function defined by using 
(\ref{sigmadef}) with Eisenstein's periodic completion  (\ref{eisfun}) as a 
replacement for the zeta function.

For any primitive
$\omega_i$,
there is a symmetric variant of the sigma function defined by
\begin{equation}
\sigma_i(z) = \sigma_i(-z) = e^{-\eta_iz}\sigma(z +
\omega_i)/\sigma(\omega_i),
\end{equation}
so $\sigma_i(\omega_i +L)$ = 0.
This is normalized so that  $\sigma_i(0)$ = 1, and
there are just three
distinct functions $\sigma_i(z)$, $i$ = 1,2,3, 
that can be labeled in a basis-dependent
way using $\omega_1$, $\omega_2$ and $\omega_3$.
The replacement of $\eta_i$ by $\widetilde \eta_i$  and $\sigma(z)$ by
$\widetilde \sigma(z)$ now
defines  modified symmetric sigma function variants $\widetilde \sigma_i(z)$.
The symmetric variants have the same quasiperiodicity as the  modified
sigma function  under \textit{even} translations:
\begin{equation}
\widetilde \sigma_i(z + 2L)
=    e^{(\pi L^*/A)(2z + L)}\widetilde\sigma_i (z).
\end{equation}

The product (\ref{full}) provides a very efficient algorithm for
numerical computation of the modified sigma function in
finite-precision floating-point arithmetic, as terms in the product eventually
become numerically equal to unity above some $n$.   An optimum
algorithm appears to involve choosing  $\omega_1$ and $\omega_2$ so
that $\pm \omega_1$, $\pm \omega_2$ and $\pm \omega_3$, are on the
boundary of the Voronoi cell of the lattice point at the origin, with
$|\omega_1|$ $\le $
$|\omega_2|$ $\le$ $|\omega_3|$.   Then use the quasiperiodicity
to reduce $z$ into this Voronoi cell, evaluate the   product  for
the reduced $z$, and use the quasiperiodicity to migrate the result
back to the original $z$.   The accuracy of the migration process can
be enhanced if it is known that an integer multiple of $z$ is a lattice
point $L$.

In the view of this author, the various  properties described above justify the view that
the modified functions  are the natural ones that
Weierstrass \textit{should} have defined,
and probably would have used,  if knowledge of the theory of quasi-modular forms had been available
to him.   Some evidence in favor of this assertion comes from an
application to the physics of electrons in a Landau level on
a two-dimensional plane that is compactified into a torus, where the
wavefunctions have a  natural representation in terms of the modified
(as opposed to the original) sigma function.

This problem can be formulated as that of non-relativistic electrons with a dispersion 
$\varepsilon(\bm p)$
= $|\bm p |^2/2m$, with dynamical momentum  $\bm p$ = $-i\hbar \bm \nabla - e\bm A(\bm
r)$,  moving on a plane though which a uniform
magnetic flux density passes.   This quantizes the kinetic energy to take discrete
values  $(\hbar^2/m\ell^2)(n + \frac{1}{2})$, $n =0,1,2,\ldots$,  where $2\pi \ell^2$ is
the area through which one London quantum $\Phi_0$ = $2\pi\hbar/e$  
of magnetic flux passes. 
In the conventional treatment using the
so-called ``symmetric gauge'', where the electromagnetic vector potential
$(A_x,A_y)$ is given by   $A_x + i A_y$ =
$i\Phi_0z/(4\pi\ell^2)$, and $z$ = $x + iy$ is a mapping from the
Euclidean plane to the complex plane, the  lowest-Landau-level ($n=0$)  wavefunctions have the form
\begin{equation}
\psi(z,z^*) =   f(z) e^{-\frac{1}{4}z^*z/\ell^2}
\end{equation} 
where $f(z)$ is a holomorphic function.

 Quasiperiodic boundary conditions under a
Bravais lattice $\Lambda$ of translations with a
primitive region of area $A$ = $2\pi N_{\Phi}\ell^2$ (and integer
$N_{\Phi}$) can be imposed
(as a technical device to avoid the problems of working with an
infinite system), so that
\begin{equation}
\psi(z+L,z^*+L^*) = \xi(L)^{N_{\Phi}} e^{K^*L - KL^*}
e^{\frac{1}{4}(L^*z-Lz^*)/\ell^2}
\psi(z,z^*) ,
\end{equation}
or
\begin{equation}
f(z+L) = e^{K^*L - KL^*} \left (\xi(L)
e^{(\pi L^*/A)(z+\frac{1}{2}L)}\right )^{N_{\Phi}}f(z).
\end{equation}
This is solved by wavefunctions with
\begin{equation}
f(z) \propto  e^{K^*z}\left (\prod_{i=1}^{N_{\Phi}}  \widetilde \sigma (z -
  w_i)\right ) , \quad  \sum_{i=1}^{N_{\Phi}} w_i =  KA/\pi.
\end{equation} 
This wavefunction has $N_{\Phi}$
zeroes in the fundamental region.
The many-particle Slater-determinant
 state with $N$ = $N_{\Phi}$ fermions that completely
fills the lowest Landau level is then given by
\begin{equation}
\Psi \propto  e^{K^*Z}\widetilde \sigma(Z -W)
\left (\prod_{i<j}\widetilde \sigma (z_i - z_j)\right )
\prod_i \left (e^{-\frac{1}{2}\pi z_i^*z_i/A}\right )^{N_{\Phi}}, \quad Z = \sum_i z_i,
\quad W = KA/\pi.
\end{equation}
Many other more-interesting model many-particle states on the torus (\textit{e.g.}, the
Laughlin state) can  be written in an  explicitly modular-invariant
way using the modified sigma function described here.   These will be
described elsewhere.

The importance of these constructions using the modified sigma
function is that they are explicitly modular-invariant, unlike most
previous representations using the Jacobi elliptic function
$\vartheta_1(u|\tau)$, $u$ = $\pi
z/2\omega_1$,
$\tau$ = $\omega_2/\omega_1$.     
When constructing model wavefunctions
on the toroidal compactification of the plane,
an important physical requirement is that they \textbf{do not depend on the
choice of basis} of $\Lambda$, \textit{i.e.}, that they are 
modular-invariant.   Use of the  modified sigma function  ensures this property is explicit.

Whether or not the simple ``modification'' described here is useful in
other applications of the Weierstrass functions, it is now clear that the
modified sigma function is the variant of choice for describing lowest
Landau-level wavefunctions.    However the simplicity of the formula (\ref{new}),
and the appearance of  ``naturalness'' of the  ``completion''  by  the inclusion of the
``missing''  $2k$ = 2 term in  the Laurent expansion (\ref{laurent}), 
suggests that \textit{all} applications of the Weierstrass functions could
benefit
from the modification proposed here, which has  antecedents in Eisenstein's work\cite{eis,weil}.

\begin{acknowledgements}
This work was supported by Department of Energy BES Grant DE-SC0002140.
\end{acknowledgements}

\end{document}